\documentclass[aps,preprint]{revtex4}%
\usepackage{amsfonts}
\usepackage{amsmath}
\usepackage[utf8]{inputenc}
\usepackage{amssymb}
\usepackage{float}
\usepackage{xcolor}
\usepackage{graphicx}%
\providecommand{\U}[1]{\protect\rule{.1in}{.1in}}

\begin{document}
\title{Exact black strings and p-branes in general relativity}

\author{Adolfo Cisterna}
\address{Universidad Central de Chile, Vicerrectoría académica, Toesca 1783, Santiago, Chile.}
\email{adolfo.cisterna@ucentral.cl}
\address{Instituto de Ciencias F\'isicas y Matem\'aticas,
Universidad Austral de Chile, Casilla 567, Valdivia, Chile.}
\author{Julio Oliva}
\address{Departamento de Física, Universidad de Concepción, Casilla, 160-C, Concepción, Chile.}
\email{julioolivazapata@gmail.com}

\begin{abstract}
\noindent We present a new set of analytic solutions of general relativity with a
negative cosmological constant, which describe black strings and black
p-branes in arbitrary number of dimensions, $D$. The solutions are supported by minimally coupled, free scalar fields, carrying finite energy density.
\end{abstract}
\maketitle

\section{Introduction} It has been well established by now that general
relativity in spacetime dimensions greater than four admits black object solutions with event horizons of different topologies \cite{Horowitz:2012nnc}; the
archetypical example being the black ring \cite{BR1, BR2}. The black ring is a Ricci flat black
hole whose event horizon has topology $S^{2}\times S^{1}$, in contrast to the $S^{3}$ topology of the Myers-Perry generalization of Kerr geometry \cite{MP}. The existence of such solutions shows how, in five or higher dimensions, the theory circumvents topological obstructions that in four dimensions it encountered for admitting hairy solutions in asymptotically flat space \cite{hawking}. 

For large
angular momentum, the black ring solution can be described by a black string
geometry, which can be thought of adding an unwrapped flat direction to the
four-dimensional Schwarzschild solution. By warping the four-dimensional Schwarzschild-AdS solution and
adding an extra dimension to it, one can easily construct analytic black strings in
AdS space. The warping factor, however, makes the AdS black string to be
non-uniform, and this introduces difficulties relative to the asymptotically flat case, specially in relation to the study of its dynamical stability as well as a proper definition of energy density. Relying on numerical tools, homogenous black strings in AdS can be constructed in pure GR with a negative cosmological constant \cite{copsey}, \cite{mannradustelea} as well as in five dimensional gauged supergravity \cite{olea}. In this paper, we prove that general relativity with negative cosmological constant, apart from admitting such warped-AdS and numerical black string solutions, also admits analytic solutions that describe homogeneous black strings and black p-branes. These solutions are supported by minimally coupled, free scalar fields and exist in arbitrary dimension $D$ greater than 4.

\section{Main Results} Consider Einstein theory in dimension $D=d+p$,
coupled to $p$ scalar fields $\psi^{\left(  i\right)  }$ with $i=1,2,...,p$.
The field equations are given by%
\begin{equation}
G_{AB}+\Lambda g_{AB}=\kappa%
{\displaystyle\sum\limits_{i=1}^{p}}
T_{AB}^{\left(  i\right)  }\ ,\label{EE}%
\end{equation}
with%
\begin{equation}
T_{AB}^{\left(  i\right)  }=\frac{1}{2}\partial_{A}\psi^{\left(  i\right)
}\partial_{B}\psi^{\left(  i\right)  }-\frac{1}{4}g_{AB}\partial_{C}%
\psi^{\left(  i\right)  }\partial^{C}\psi^{\left(  i\right)  }\ ,
\end{equation}
and%
\begin{equation}
\square\psi^{\left(  i\right)  }=0\text{ with }i=1,2,...p\ .
\end{equation}
Here, $G_{AB}$ is the Einstein tensor. Hereafter, we will set $\kappa=16\pi G=1$.

\noindent The theory defined above admits the following solution%
\begin{equation}
ds^{2}  =-F(r) dt^2+\frac{dr^2}{F(r)}+r^2d\Omega_{d-2,\gamma}^{2}+\delta_{ij}dx^{i}dx^{j}\label{themetric}%
\end{equation}
provided
\begin{equation}
F(r)=  \gamma-\frac{2\mu}{r^{d-3}}-\frac{2\Lambda r^{2}}{\left(
d-1\right)  \left(  d+p-2\right)  }\ ,\label{MU}
\end{equation}
with $x^{i}$ ($i=1,...,p$) being Cartesian coordinates. These are the coordinates along the flat $p-$brane. Remarkably, the solution for the fields takes the simple form%
\begin{equation}
\psi^{\left(  i\right)  }=\lambda x^{i}\ ,\label{axions}%
\end{equation}
with%
\begin{equation}
\lambda^{2}=-\frac{4\Lambda}{\left(  d+p-2\right)  }\ .\label{lamcua}%
\end{equation}
That is, the scalar fields have a linear dependence with the coordinates $x^i$. In (\ref{themetric})-(\ref{MU}), $\mu$ appears as an arbitrary integration constant, and $\gamma=\pm1,0$ is the
curvature of an Euclidean manifold of constant curvature of dimension $d-2$ and line element
$d\Omega^2$. Note that (\ref{lamcua}) forces the {\it bare} cosmological constant
$\Lambda$ to take a negative value. 

The solution presented above is the first of its type of having been found analytically; namely, it is the first homogeneous, analytic black
$p$-brane solution of Einstein equations with non-vanishing cosmological constant.

Spacetime (\ref{themetric}) is asymptotically $AdS_{d}\times R^{p},$ with
the curvature radius of the $AdS_{d}$ factor given by%
\begin{equation}\label{ldressed}
\frac{1}{l^{2}}=-\frac{2\Lambda}{\left(  d-1\right)  \left(  d+p-2\right)
}=-\frac{2\Lambda}{\left(  D-p-1\right)  \left(  D-2\right)  }\ .
\end{equation}

Notice that this value for the {\it dressed} $AdS_{d}$ curvature radius $l$, obtained from (\ref{ldressed}), differs from the {\it bared} value of the maximally symmetric AdS$_{D}$ solution of the theory, $l_{0}^{-2}=-2\Lambda/\left[  \left(  D-1\right)  \left(
D-2\right)  \right]  $. In general, $l\leq l_{0}$, with the upper bound corresponding to $p=0$.

We can choose $\gamma=\pm1,0$, which leads to three possible local geometries for the asymptotic boundary. That is to say, the holographic dual field theory can in principle be formulated either on $R^{D-1}$ for $\gamma=0$, $R\times S^{d-1}\times R^{p}$ for $\gamma=1$, or $R\times H^{d-1}\times R^{p}$ for $k=-1$.

\bigskip

\section{General construction} Let us now consider a general $D$-dimensional metric of the
form%
\begin{equation}
ds_{D}^{2}=d\tilde{s}_{d}^{2}+\delta_{ij}dx^{i}dx^{j}\ ,
\end{equation}
and the set of scalar fields $\psi^{\left(  i\right)  }=\lambda x^{i}$, where we
have split the indices in such a way that Greek indices and tilded objects live on
the manifold with line element $d\tilde{s}$, while lowercase Latin indices run
along the $p$ extended directions. 

Einstein equations (\ref{EE}) projected along
the manifold $d\tilde{s}$ and the extended directions $x^{i}$, respectively
reduce to%
\begin{equation}
\tilde{G}_{\mu\nu}+\left(  \Lambda+\frac{p\lambda^{2}}{4}\right)  \tilde
{g}_{\mu\nu}=0\ ,\label{gen4}%
\end{equation}
and%
\begin{equation}
\tilde{R}=2\Lambda-\left(  1-\frac{p}{2}\right)  \lambda^{2}\ .\label{genx}%
\end{equation}

The compatibility of the trace of (\ref{gen4}) --obtained by contracting such
equation with $\tilde{g}^{\mu\nu}$-- with equation (\ref{genx}) implies that
the constant $\lambda$ must be fixed as in (\ref{lamcua}). In other terms, the
configuration of the scalar fields induces a shift in the
cosmological constant of any $d$-dimensional Einstein manifold. Therefore, on the transverse section of the $p$-brane we can consider any
solution to Einstein equation in $d$ dimensions, provided (\ref{gen4}) is obeyed. We can, for
example, consider the asymptotically AdS rotating solution of general
relativity with negative cosmological constants, which is characterized by its mass as well as
$\left[  \frac{d-2}{2}\right]  $ angular momenta \cite{hawkingrot, gibbonslupagepope1, gibbonslupagepope2}, to construct black
strings in $AdS_{d}\times R^{p},$ with a rotating black hole on the brane.

\section{Further comments} The black $p$-brane solutions are supported by the
scalar fields $\psi^{\left(  i\right)  }$, which are linear on the coordinates
$x^{i}$. Even though these fields diverge in the limit $x^{i}\rightarrow\pm\infty$, they yield finite energy density. That is, the divergence merely comes from the non-compactness of the extended directions. In fact, the $tt$ component of the energy-momentum tensor for the collection of
$\psi^{\left(  i\right)  }$ turns out to be independent of the coordinates $x^{i}$. Therefore, one can properly define the energy density, as in the Ricci-flat,
homogenous black strings. 

Solution (\ref{themetric}%
)-(\ref{axions}) exists due to the fact that, despite the metric being
homogenous, the scalar fields break translational symmetry. This idea has been used in many
different contexts, for example in the construction of boson stars and other gravitational solitons \cite{Liebling:2012fv, canforamaeda} as well as for rotating hairy black holes  \cite{herdeiro}.

Each of the scalars in the solution can be dualized to $\left(  D-1\right)  $-forms. Both the formulations in terms of scalars and in terms of the $\left(
D-1\right)  $-forms have resulted very useful to
construct asymptotically AdS, planar black holes in the presence of other
matter fields, which is of great importance in the context of AdS/CFT
correspondence and, in particular, to its applications to condensed matter (see e.g. \cite{Bardoux:2012tr, Andrade:2013gsa}).

The entropy density is given by the area law, namely%
\begin{equation}
s=\frac{S}{V}=\frac{r_{+}^{d-2}\sigma}{4G}=4\pi r_{+}^{d-2}\sigma\ ,
\end{equation}
where $\sigma$ is the unit volume of $\Omega_{d-2,\gamma}$, $V$ is the volume
of the extended directions and $r_{+}$ is the largest root of the equation $0=F\left(
r_+\right)  :=-g_{tt}$. The temperature, $T$, can be
obtained as usual from the period of the Euclidean time that yields a
regular Euclidean section of the imaginary continuation $t\to i t$. This yields,
\begin{equation}
T=\frac{f^{\prime}\left(  r_{+}\right)  }{4\pi}=\frac{1}{4\pi}\left(
\frac{\left(  d-3\right)  }{r_{+}}-\frac{2\Lambda}{d+p-2}r_{+}\right).
\end{equation}

Since there is no charge associated to the free scalar fields in this family of solutions, the mass $M$
--or the energy density $m=M/V$-- can directly be obtained from the first
law of black hole mechanics; namely $dm=Tds$. Therefore,
\begin{equation}
m=\sigma\left(  d-2\right)  r_{+}^{d-3}\left(  1-\frac{2\Lambda}{\left(
d-1\right)  \left(  d+p-2\right)  }r_{+}^{2}\right)  \ .
\end{equation}

Before concluding, let us mention that black strings suffer from Gregory-Laflamme (GL) instabilities \cite{GL}, namely long-wavelength, perturbative instability triggered by a mode travelling along
the extended directions. This kind of instability goes beyond the realm of
general relativity, and pervades also black string solutions in other theories, like higher-curvature
gravities \cite{Brihaye:2010me, terance, ubs1, ubs2, Chen:2017rxa}. Numerical simulations show that in five dimensions the GL
instability leads to the formation of a naked singularity \cite{choptuikfinalstage, LPfinalstage}, while
thermodynamical arguments indicate that for dimensions greater than 13 the
final stage of the instability could be an inhomogeneous black string \cite{Sorkin}. The latter
has been recently confirmed in the large $D$ limit \cite{Emparan:2015gva}. It is likely that
small (as compared with $l$) black strings will suffer from a GL instability. Whether or not that is the case, is beyond the scope of the present work and is matter of our current research. One
might nevertheless expect, in the context of the AdS/CFT duality \cite{adscftmaldacena}-\cite{adscftwitten}, that if the dual CFT has a well defined evolution,
the existence of an instability shouldn't lead to a naked singularity in the bulk as in
the Ricci flat case.

\section{Acknowledgement}
We thank Gaston Giribet for many enlightening comments. A.C. work is supported by FONDECYT project N\textordmasculine3150157 and Proyecto Interno Ucen I+D-2016, CIP2016. We appreciate the support of the Associate Program of the International Center for Theoretical Physics, ICTP, where this work was done.

\end{document}